\documentclass[twocolumn,showpacs,preprintnumbers,amsmath,amssymb]{revtex4}

\usepackage{graphicx}
\usepackage{epsfig}

\usepackage{longtable}
\usepackage{longtable}

\begin{document}

\title{Hindered alpha decays of heaviest high-K isomers.}

\author{P.~Jachimowicz}
 \affiliation{Institute of Physics,
University of Zielona G\'{o}ra, Szafrana 4a, 65516 Zielona
G\'{o}ra, Poland}

\author{M.~Kowal} \email{michal.kowal@ncbj.gov.pl}
\affiliation{National Centre for Nuclear Research, Ho\.za 69,
PL-00-681 Warsaw, Poland}

\author{J.~Skalski}
\affiliation{National Centre for Nuclear Research, Ho\.za 69,
PL-00-681 Warsaw, Poland}

\date{\today}

\begin{abstract}

  To find candidates for long-lived high-K isomers in even-even Z=106-112
 superheavy nuclei we study dominant alpha-decay channel of two- and four-quasi-particle
 configurations at a low excitation. Energies are calculated within the
 microscopic - macroscopic approach with the deformed Woods-Saxon potential.
 Configurations are fixed by a standard blocking procedure and their energy
  found by a subsequent minimization over deformations.
  Different excitation energies of a high-K configuration in parent and
  daughter nucleus seem particularly important for a hindrance of
  the alpha-decay. A strong hindrance is found for some four-quasi-particle
  states, particularly $K^{\pi} = 20^{+}$ and/or $19^{+}$ states in
  $^{264-270}$Ds. Contrary to what was suggested in experimental papers,
  it is rather a proton configuration that leads to this strong hindrance.
  If not shortened by the electromagnetic decay, alpha half-lives of
 $\sim$ 1 s could open new possibilities for studies of chemical/atomic
  properties of related elements.

\end{abstract}

\pacs{21.10.-k, 21.60.-n, 27.90.+b}

\maketitle

\section{INTRODUCTION}

 Superheavy elements are highly unstable systems with extremely low production
 cross sections. As the creation of new ones is very difficult, as a parallel
 or additional line of study one could try a search for new, long-lived
 metastable states of already known nuclei.
 It is well known that an enhanced stability may result from the K-isomerism
  phenomenon \cite{Walker,Xu} which is based mainly on the (partial)
 conservation of the K-quantum number \footnote{K ($\Omega$ for a s.p.
 state) is a total angular momentum along the symmetry axis and is a "good"
 quantum number in the case of axial symmetry.}.
 Low-lying high-K configurations occur when high-$\Omega$ orbitals lie
 close to the Fermi energy. When such orbitals are intruders, the resulting
 unique configuration may have even longer half-life.
 Possible K values grow with larger-$j$ subshells becoming occupied, that
 means for larger $Z$ and $N$.

 Currently known are many rotational bands build on 2 q.p. or 4 q.p. K-isomeric
 band-heads\cite{Hof,Her,Tan,Kon,Hes,Cla}.
 The structure of expected long-lived multi-quasiparticle high-spin isomers in
 some even-even SH nuclei was analyzed e.g. in \cite{Xu}. In particular,
 the assignment of 9$^{-}$ or 10$^{-}$
 two quasineutron configurations for the $6.0 ^{+8.2}_{-2.2}$ms isomer in
 $^{270}$Ds (the heaviest isomer known) was proposed \cite{Hof}.
 Let us stress that the half-life of this isomer is much longer
 than that of the ground state ($100 ^{+140}_{-40}$ $\mu$s). The same holds
 for the 8$^{+}$ isomer in $^{256}$Es, with the half-life of 7.6 h -
 significantly longer than 25 min of the g.s. Another interesting example
 is a 16$^{+}$ or 14$^{+}$ state in $^{254}$No, with a half-life of 184 $\mu$s,
  at 2.93 MeV above the g.s \cite{Her,Tan,Kon,Hes,Cla}.
  Other, four-qp isomers in nuclei around $^{254}$No were postulated in
  \cite{Liu1}.
 The current experimental knowledge on isomers in the heaviest nuclei can be
 found in \cite{Ack,Asai,Drac}, while theoretical overview based on the
 Nilsson - Strutinsky approach was given in \cite{walkerPS}.
 Let us emphasize that all K-isomeric states described above are related to
 typical prolate equilibria.
  A quite new possibility for high-K isomers in the superdeformed oblate minima
  in some SH nuclei was indicated in \cite{SDO}. We also would like to note
 that both measurements and predictions are performed mainly for
  even-even nuclei. Quite recently, we have also predicted high-K ground state
  configurations in odd and odd-odd systems \cite{Kisomers}.
 In this case, a particular situation occurs above double-closed subshells:
 $N=162$ and $Z=108$, where two intruder orbitals: neutron $13/2^-$ from
$j_{15/2}$ and proton $11/2^+$ from $i_{13/2}$ spherical subshells are
 predicted. These orbitals combine to the $12^-$ g.s. in $Z=109$, $N=163$.

 Although the existence of isomeric states is rather well established,
 hindrance mechanisms responsible for slowing down their radioactive decay
  are still poorly understood.
 In particular, a hindrance of alpha decay, which is the main decay channel
  in superheavy nuclei with $ Z=106\div118$, is not well elucidated.
  For those nuclei, hindered $\alpha$ transitions would lead to an increase in
 $\alpha$-lifetimes and extra stability.
 This is why, in this letter, we are going to estimate effects protecting
 high-K nuclear states against this decay mode.
Our ultimate goal is to find candidates for long-lived nuclear configurations.

\section{THE METHOD}

Predictions for high-K multi-quasiparticle nuclear configurations require
 a model that satisfactorily describes well-known basic nuclear properties as:
 ground state masses, fission barriers, equilibrium deformations etc.
 Important is the existence of sufficiently distinct energetic shell gaps: two
 of them in the proton spectrum, at around Z=100 and Z=108, and next two at
 N=152 and N=162 in the neutron spectrum.
 Without this, the spectroscopic studies (especially devoted to high-K-isomers)
 are just unfeasible in this region of nuclei.
 The only model available which satisfies both conditions simultaneously is
 the Microscopic-Macroscopic (MM) approach.
 Our model is based on the Yukawa-plus-exponential energy in the macroscopic
 part \cite{KN}, while the single-particle spectrum is obtained from
 a deformed Woods-Saxon (WS) potential \cite{WS}, diagonalized in the
 deformed harmonic oscillator basis.
 Within this model, with parameters adjusted
 to heavy nuclei in \cite{WSpar}, it was possible to reproduce data on
 first \cite{Kow},  second \cite{kowskal,IIbarriers} and third
 \cite{IIIbarriers1,IIIbarriers2} fission barriers in actinides,
 systematically predict ground states \cite{archive,jach2017}, $Q_{\alpha}$
 values \cite{Jach2014} and saddle-points \cite{jachbar2017} in even- and
 odd-Z/N superheavy nuclei up to $Z=126$ and gain some insight in
 a SH region beyond $Z=126$ \cite{BroSkal}.

  The blocking method induces a too large reduction in the pairing gap
  for multi-quasiparticle states. This causes an underestimate in their
  excitation energies. However, the effects we are going to discuss here are
  independent of this deficiency: 1) the considered configurations come
  out as the lowest ones in any method - see e.g. \cite{Liu1}, so they are
  anyway the main candidates for isomers, 2) within the accuracy
  of the current models, even more precisely calculated excitation energies
  cannot prove/disprove an isomeric character of a many-quasiparticle state,
  3) the crucial effect, the reduction of energy of $\alpha$-transistions
  between the same configurations in parent and daughter, is in a large part
   independent of the pairing reduction by blocking - the blocking effect
   in excitation energies in both parent and daughter mostly cancels.
  Therefore we decided to present the results obtained without any specially
   adjusted parameters.

The choice of a proper deformation space is important.
Recently, an effect of the deformation $\beta_{60}$ on high-K isomer properties in superheavy nuclei has been discussed by Liu et. al. \cite{Liu2011}.
The authors noted a wider shell gap around Z=100 and N=150 after the inclusion
 of $\beta_{60}$, which gave much better agreement with existing experimental
 data.
 This conclusion is compatible with the previous one, given by Patyk and
 Sobiczewski in \cite{Patyk1,Patyk2}.
 In this work, the shape of a nucleus is parameterized via spherical
 harmonics ${\rm Y}_{lm}(\vartheta ,\varphi)$ in a four-dimensional
 deformation space:
$R(\vartheta ,\varphi)= c(\{\beta\}) R_0 \{1 +\beta_{20}{\rm Y}_{20}+\beta_{40}{\rm Y}_{40}+\beta_{60}{\rm Y}_{60}+\beta_{80}{\rm Y}_{80}\}$,
where $c(\{\beta\})$ is the volume-fixing factor and $R_0$ is the radius of a
 spherical nucleus.
One can see that the used parameterization still contains one more dimension
 compared to \cite{Liu2011}.
In the considered region, $106\leq Z \leq 118$, nuclei are expected to be
 reflection-symmetric in the ground- and excited states, so
 we do not need odd multipole deformations here and
 the intrinsic parity of states is well-defined.
Mother, as well as daughter nuclei, are also assumed axially-symmetric,
 what gives K as a good quantum number (for a search of more exotic nuclear
 shapes in SH nuclei see \cite{jach2017}). Admittedly, this assumption cannot
 be exact for high-K states, in which the time-reversal
 breaking effects are expected to break axial symmetry to some degree.
 To obtain excitation energies, after blocking of a chosen configuration,
 a four-dimensional minimization over $\beta_{20}-\beta_{80}$ was performed.
\begin{figure}[h]
\centerline{\includegraphics[scale=0.85]{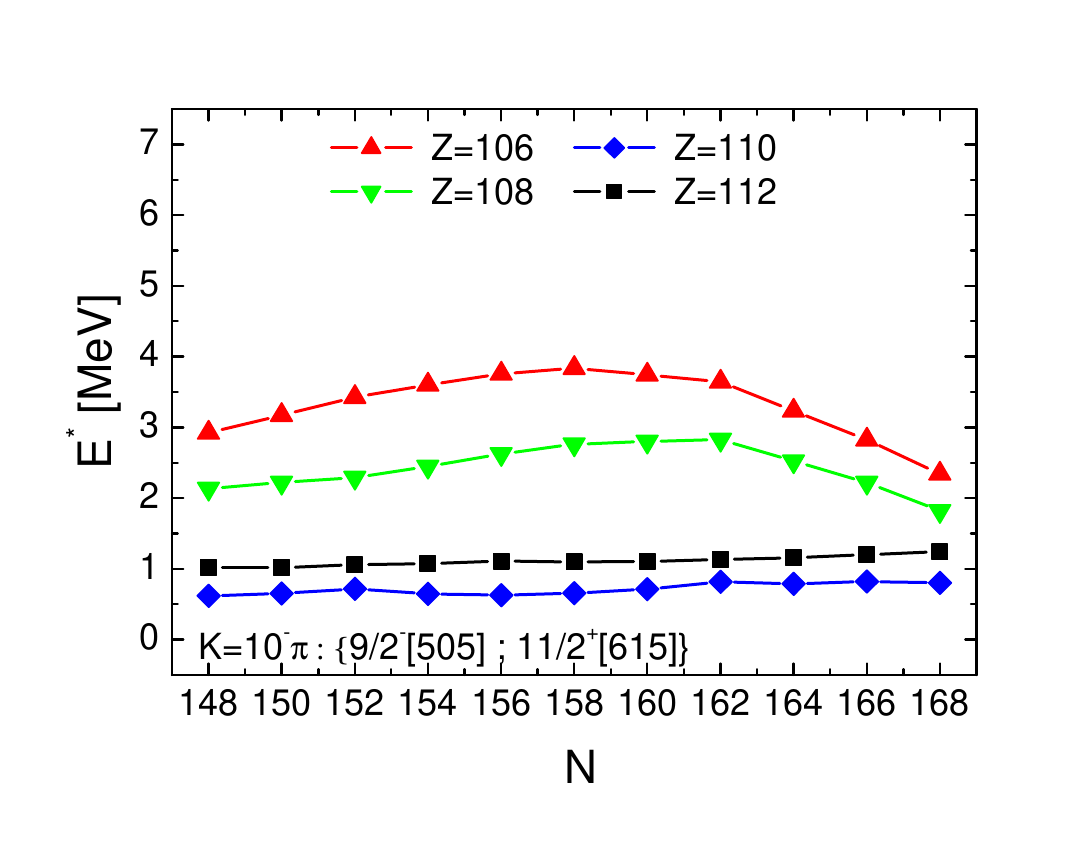}}
 \caption{{\protect Excitation energy of the two-proton configuration in
  the isotopic chains for Sg, Hs, Ds and Cn.}}
\label{10-p}
\end{figure}

\section{RESULTS AND DISCUSSION}

We will consider here three characteristic multi-quasiparticle configurations,
 namely:
i) two-neutron (2 q.p.):
$ K^{\pi} = 10^{-} \nu : \{9/2^{+}[615], 11/2^{-}[725] \}$;
ii) two-proton (2 q.p.):
$ K^{\pi} =10^{-} \pi : \{(9/2^{-}[505], 11/2^{+}[615]\}$;
iii) two-proton - two-neutron (4 q.p.):
$ K^{\pi} =20^{+} \nu \pi:  ( 10^{-} \nu : \{(9/2^{+}[615], 11/2^{-}[725]\} \otimes 10^{-} \pi : \{(9/2^{-}[505], 11/2^{+}[615]\})$.
 The results obtained for another low-lying configuration, two-neutron (2 q.p.)
$ K^{\pi} = 9^{-} \nu : \{7/2^{+}[613], 11/2^{-}[725] \}$, considered
 e.g. in \cite{Liu1}, are similar to those for the above configuration i).

 We begin our analysis with excitation energies in isotopic chains.
 Energies of two-proton quasiparticle states are shown in Fig. \ref{10-p}.
 As can be seen, those excited configurations are very low-lying for Ds and Cn
 nuclei what makes them promising candidates for (2 q.p.) isomers.
One can also see that the lowest energies are obtained in Darmstadtium and
 that in both, Ds and Cn nuclei, excitation energies show very weak isotopic
 dependence.

\begin{figure}[h]
\centerline{\includegraphics[scale=0.85]{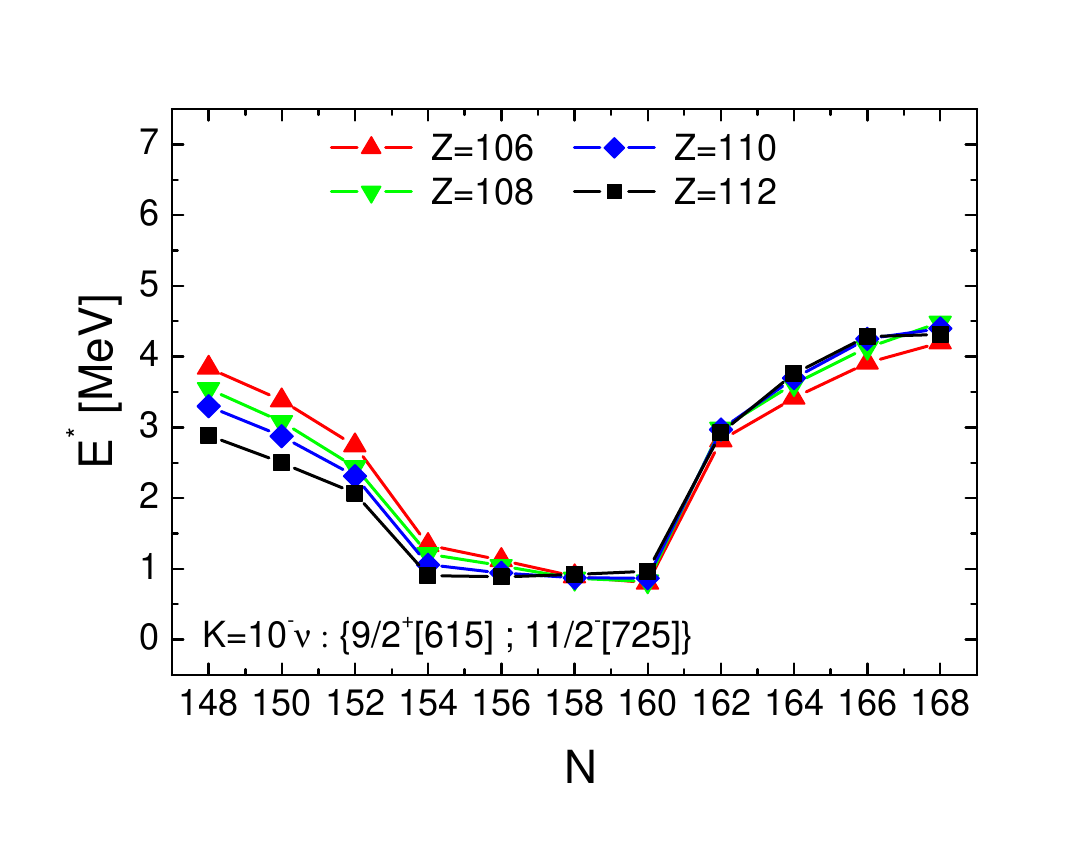}}
 \caption{{\protect The same as in Fig. \ref{10-p}, but for the two-neutron
 configuration.}}
\label{10-}
\end{figure}

\begin{figure}[h]
\centerline{\includegraphics[scale=0.85]{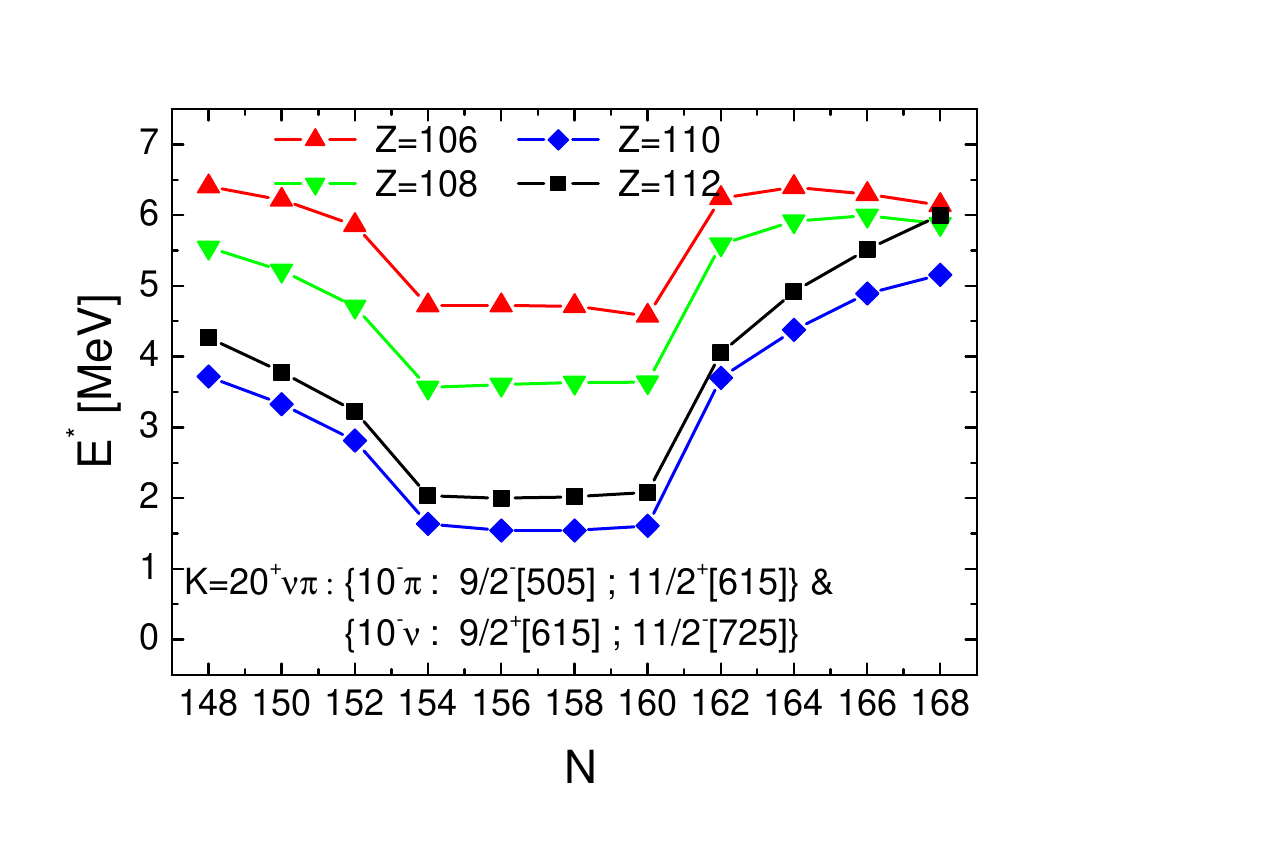}}
 \caption{{\protect The same as in Fig. \ref{10-p}, but for the two-neutron
 plus two-proton configuration.}}
\label{20+}
\end{figure}

 From Fig. \ref{10-} one can choose the best candidates for two quasi-neutron
 isomers. Among all considered nuclei, the lowest energies occur for
 $N=154-160$.
 Figures 1 \& 2 allow a prediction of the most favorable
  four quasi-particle isomers.
 As can be seen in Fig. \ref{20+}, 4 q.p. high-K isomeric states can appear
 most likely in $^{264-270}$Ds and $^{266-272}$Cn because of the smallest
 excitation energy. In particular, this energy amounts to 1.4 MeV, which
 means that it is the sum of excitations of the individual s.p. states,
 with nearly vanishing pairing gap. It is much less than 2.41 MeV given
 by Liu \cite{Liu1} et. al., who used the more realistic Lipkin-Nogami
 procedure.
\begin{table}
\caption{ Calculated  decimal logarithms of various hindrance factors
 for 2 q.p. neutron $ K^{\pi} = 10^{-} \nu : \{9/2^{+}[615], 11/2^{-}[725] \}$
 and proton $ K^{\pi} =10^{-} \pi : \{(9/2^{-}[505], 11/2^{+}[615]\}$
 configurations in $^{270}$Ds:
  $Log_{10}[HF]_{Q}$ related to the $Q_{\alpha}$ change;
 $Log_{10}[HF]_{L}$ related to the angular momentum change (calculated within
 the WKB aproximation \cite{dub}); $Log_{10}[HF]_{S}$ related to the structure
  change, taken from \cite{Delion}. The experimental $Log_{10}(T_{1/2})$ for
  the g.s. is given in parenthesis.
\label{table1}}
\begin{tabular}{|c|c|c|c|c|}
 \hline
   $K^{\pi} = 10^{-} \nu$     &   $gs \rightarrowtail gs $    &    $ex \rightarrowtail ex $  &    $ex \rightarrowtail gs $ & $gs \rightarrowtail ex$   \\
 \hline
      $Q_{\alpha}$                                             & 11.38 &  11.38  &   12.25  &   10.51        \\
 \hline
      $Log_{10}[HF]_{Q}$                                       & 0      & 0       &   -1.82  &   2.07     \\
  \hline
      $Log_{10}[HF]_{L}$                                       & 0      & 0       &   4.06  &   4.17     \\
   \hline
      $Log_{10}[HF]_{S}$                                       & 0      & 0       &   4.74  &   4.74     \\
   \hline
    $Log_{10}[HF]_{TOT}$                                       & 0      & 0       &   6.98   &   10.98     \\
   \hline
      $Log_{10} [T_{1/2}(s)]$                                  & -4.46(-3.69)     &   -4.46  &   2.52 &   6.41       \\
   \hline
   $K^{\pi} =10^{-} \pi$     &   $gs \rightarrowtail gs $    &    $ex \rightarrowtail ex $ &    $ex \rightarrowtail gs $ & $gs \rightarrowtail ex$   \\
 \hline
      $Q_{\alpha}$                                             & 11.38  &  9.33  &   12.09 &   8.62        \\
 \hline
    $Log_{10}[HF]_{Q}$                                         & 0      & 5.44   &   -1.50  &   8.00     \\
  \hline
      $Log_{10}[HF]_{L}$                                       & 0      & 0      &   4.16  &   4.67     \\
   \hline
      $Log_{10}[HF]_{S}$                                       & 0      & 0      &   4.08  &   4.08     \\
   \hline
      $Log_{10}[HF]_{TOT}$                                     & 0      & 5.44   &   6.74  &   16.75     \\
   \hline
      $Log_{10} [T_{1/2}(s)]$                                  & -4.46(-3.69)     &   0.98  &   2.28 &   12.29       \\
   \hline
\end{tabular}
\end{table}

 Stability of high-spin isomers against alpha decay is determined mainly by
 three factors:
i) the overlap between final and initial states wherein a similar structure of
 states favors the transition between them;
ii) a change in angular momentum - a significant change is associated with a
 large centrifugal barrier which blocks a decay;
iii) transition energy, which we shall also call $Q_{\alpha}$ for a
  given decay, that follows from the $Q_{\alpha}$ value for the
 g.s.$\rightarrow$g.s. transition and the difference in the excitation
 energies of the initial and final state in, respectively, mother and daughter
  nucleus. We intend to analyze each of them.

 For this purpose, we define a hindrance factor as a ratio of half-lives
 for two transitions: between the excited states
 with the same fixed structure and between the ground states:
 $HF=[T^{ex \rightarrowtail ex}_{1/2}/T^{gs \rightarrowtail gs}_{1/2}]$.

 Appropriate hindrance factors are collected in Table I for two-neutron and
 two-proton configurations:
$ K^{\pi} = 10^{-} \nu : \{9/2^{+}[615], 11/2^{-}[725] \}$; $ K^{\pi} =10^{-} \pi : \{(9/2^{-}[505], 11/2^{+}[615]\}$;.

 The influence of the centrifugal term can be estimated by evaluating the
 WKB integral.
 Since for a proton and neutron configuration a change in the angular momentum
 $\Delta L=10\hbar$ is the same, the centrifugal barrier
  $V_{L}= L(L+1)/(2 J_{x}) = 1.2$ MeV increases the half-life of both
 by about four orders of magnitude. Moment of inertia can be found in \cite{Hi}
 of about $J_{x} = 47 \hbar^{2}/MeV$.
An insignificant difference between them results from a slightly different
 turning points, calculated for slightly different  $Q_{\alpha}$.

 Structural hindrance factors $Log_{10}[HF]_{S}$ have been calculated by
 Delion et. al. in \cite{Delion}.
 For our nonaligned states, the authors of \cite{Delion} gave:
 $Log_{10}[HF]_{S} = 4.74 $ and $4.08$, for $ K^{\pi} = 10^{-}\nu$ and
 $K^{\pi} =10^{-} \pi$, respectively
 \footnote{These authors' definition of the HF is, from a formal point of view,
  different from ours:
 "Hindrance factor is a measure of whether the parent nucleus would prefer to
 decay from the exited state instead of the ground state, or perhaps more
 realistically to which extent the parent nucleus would prefer to decay from
  the g.s."}.
  So for both proton and neutron excited states the size of the structural
  hindrance is again similar.

 Hindrances corresponding to changes in $Q_{\alpha}$ ($\Delta Q_{\alpha} = Q^{ex \rightarrowtail ex}_{\alpha}-Q^{gs \rightarrowtail gs}_{\alpha} $)
  are completely different though.
While for $K^{\pi} = 10^{-}\nu$ there is no hindrance (adequate hindrance
 factor is zero),
for $K^{\pi} =10^{-} \pi$ quite a strong hindrance is calculated -
 $Log_{10}[HF]_{Q} = 5.44$.
This fact is connected with an opposite excitation scheme and follows from
 the deformed single-particle Woods-Saxon spectrum.
 In $^{270}$Ds, the neutron state: $11/2^{-}[725]$ lies $\simeq 0.2$ MeV below
 the neutron Fermi level, while $9/2^{-}[734]$ lies only
 $\simeq 0.5$ MeV deeper. This gives the excitation energy of about 0.7 MeV,
 visible in Fig.\ref{10-p}. As mentioned in the introduction,
  this two-neutron configuration,
 $K^{\pi} = 10^{-} \nu \{\{9/2^{+}[615], 11/2^{-}[725] \}$,
  was assigned in \cite{Hof} to the 1.13 MeV isomer in $^{270}$Ds.
 In protons, the $11/2^{+}$, dominantly $[615]$ state, lies at the Fermi level.
 Thus the excitation of the two-proton configuration equals roughly the
 excitation of the $9/2^{-}[734]$ state, $\simeq 0.5$ MeV below the Fermi
 level, see Fig. \ref{10-}.
 So, the excitation energies of the proton and neutron configurations in
 the parent nucleus are similar.
 In the daughter, the energy of the two quasi-proton excitation is nearly 5
 times larger ($\simeq 2.5$ MeV) than that of the two quasi-neutron one.
 As a result, the same two-proton configuration lies at much higher energy
 in the daughter nucleus.
 This leads to a significant reduction in the energy of the alpha transition
 and an increase in the $\alpha$ half-life.

 A very similar scheme of excitations occurs also in $^{268}$Ds and $^{266}$Ds.
 A slightly different situation one can observe in the alpha decay of
 $^{264}$Ds.
 The $K^{\pi} = 20^{+}$ ($19^{+}$) state in $^{260}$Hs has a sizable
  excitation built to an equal degree by a two quasi-proton and two
 quasi-neutron components.
  This leads to a reduction of the transition energy and a greater stability
 of this state.

  It should be emphasized that in our model there is no reason
  for a hindrance of the alpha decay between the same neutron
 $K^{\pi} = 10^{-} \nu$ configurations: there is no spin or structure change,
  and no change in $Q_{\alpha}$ with respect to the g.s.
  However, a transition from this state to the
 ground state is significantly hindered ($HF_{TOT} \simeq 10^{7}$).
 The alpha decay from a two quasi-proton state has a hindrance factor
  $HF_{TOT} = 10^{5.44}$ for the configuration-preserving transitions
 ($ex \rightarrowtail ex $) and $HF_{TOT} \simeq 10^{7}$ for the
 $ex \rightarrowtail gs $ transitions.

  As follows from the above estimates, crucial is the hindrance in the
 fastest channel, between two identical configurations.
 This is especially true for four quasi-particle states, for which one expects
 a significant increase in the centrifugal barrier. With $\Delta L=20\hbar$, the centrifugal barrier $V_{L}=4.5$ MeV
  gives a huge $HF_{L}=10^{12}$.
 A similar magnitude of the centrifugal barrier effect, $\simeq 10^{13}$,
  was estimated by Karamian et. al. \cite{Karamian} for the $\alpha$
 decay of the 4 q.p. $16^+$ isomer in $^{178}$Hf, the $^{178}Hf ^{m2}$ state.

 A structural hindrance for 4 q.p. isomers is also substantial. If one assumes
 that it is a product of the hindrance factors for protons and neutrons, one
 obtains $HF_{L}=10^{9}$.
 Taken together, this leads to the conclusion that transitions
 $ex \rightarrowtail gs $ or $gs \rightarrowtail ex $ are excluded.
 Therefore, we concentrate in a further discussion (in Fig. \ref{Q110},
 \ref{T110}, \ref{HF110}, \ref{HF112}) on configuration-preserving transitions:
  $g.s.\rightarrowtail g.s.$, $10^{-}_{\pi} \rightarrowtail 10^{-}_{\pi}$,
 $10^{-}_{\nu} \rightarrowtail 10^{-}_{\nu}$
 and $20^{+}_{\pi \nu} \rightarrowtail 20^{+}_{\pi \nu}$.
 In such transitions, a hindrance follows from differences in alpha-decay
 energy $Q_{\alpha}$.

Let us examine the properties of 4 q.p. states in two isobaric nuclei
 $^{270}$Cn and $^{270}$Ds.
\begin{table}
\caption{  $Q_{\alpha}$-values (in MeV) and hindrance factors corresponding to
 the change $\Delta Q_{\alpha} = Q^{ex \rightarrowtail ex}_{\alpha}-Q^{gs \rightarrowtail gs}_{\alpha} $
 for the $ K^{\pi} =20^{+} \nu \pi: ( 10^{-} \nu : \{(9/2^{+}[615],
 11/2^{-}[725]\} \otimes 10^{-} \pi : \{(9/2^{-}[505], 11/2^{+}[615]\})$
 configuration in $^{270}$Cn and $^{270}$Ds, calculated using: WKB method (WKB)
  \cite{dub}, the formula of Royer \cite{Roy} (ROY), and the Viola-Seaborg-type
 formula by Parkhomenko and Sobiczewski (PS) \cite{park}.
\label{table1}}
\begin{tabular}{|c|c|c|c|c|c|}
 \hline
        &   $Q_{\alpha}$ &   $\Delta Q_{\alpha}$     &   $Log^{WKB} [HF]$ &   $Log^{ROY}[HF]$ &   $Log^{PS} [HF]$       \\
 \hline
      $^{270}$Cn                                        & 13.06 &   0.48  &   -0.87  &  -0.92 &  -0.88    \\
 \hline
      $^{270}$Ds                                        & 9.36  &  -2.02  &   6.75   &  5.42  &  5.13     \\
 \hline
\end{tabular}
\end{table}
\begin{figure}[h]
\centerline{\includegraphics[scale=0.45]{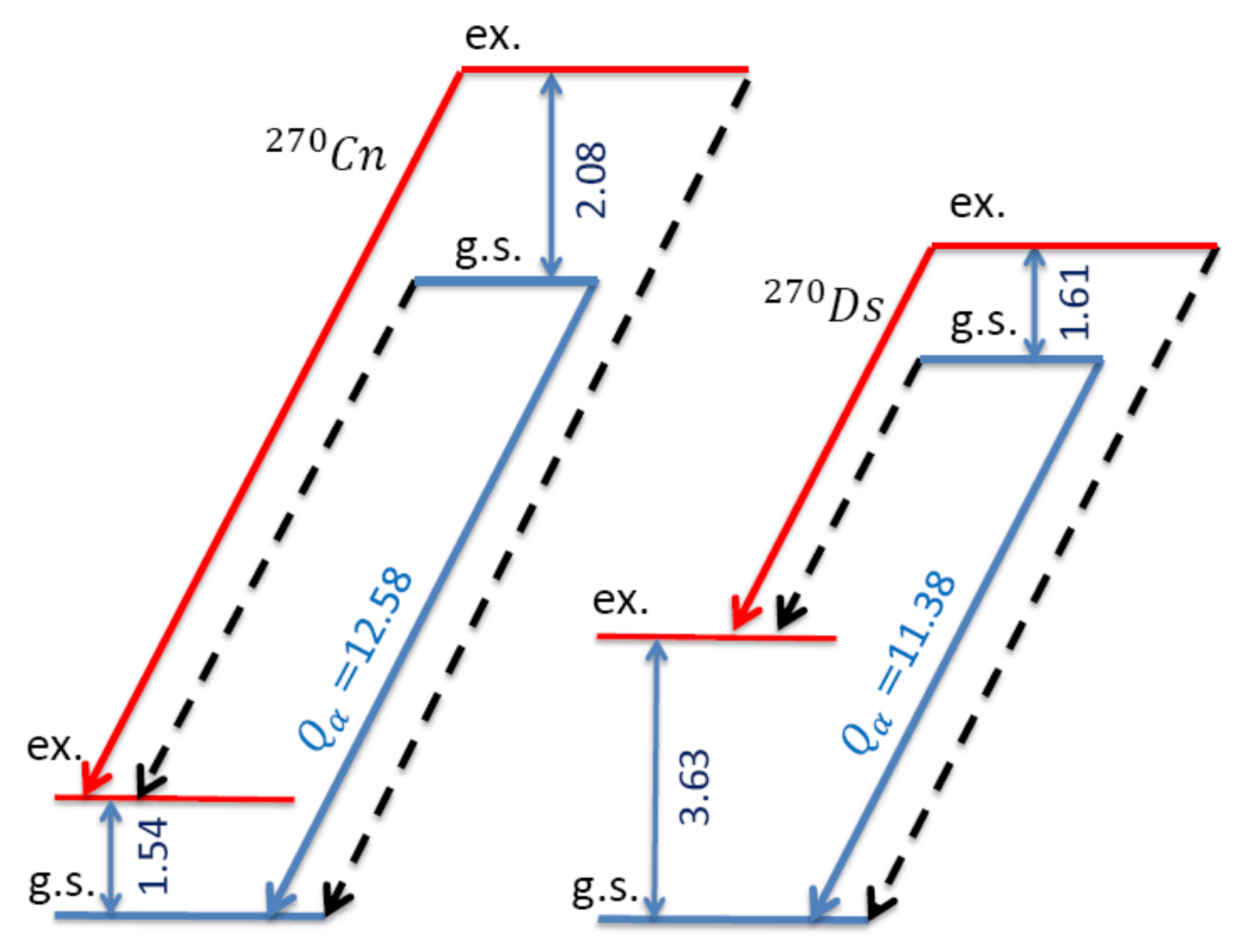}}
 \caption{{\protect Decay scheme of $^{270}$Cn and $^{270}$Ds shown
 relative to the daughter g.s.}}
\label{rysschem}
\end{figure}
Due to a difference in excitation energy in the mother and daughter,
 in $^{270}$Ds (the right part of Fig. \ref{rysschem})
 the $\alpha$-transition energy is much smaller than between the ground states.
 Thus, it seems that the $\alpha$-decay will be correspondingly slower.
 In the neighbouring $^{270}$Cn, an opposite energy relation will make
 the 4 q.p.$\rightarrowtail$ 4 q.p. transition about two orders of magnitude faster
  than the  $gs \rightarrowtail gs $ one.
 In both cases, 2 q.p. proton excitations are responsible for this scenario,
 see Fig. \ref{10-p}.
 Hindrance factors for decays of $ K^{\pi} =20^{+} \nu \pi$ states in
 $^{270}$Ds and $^{270}$Cn are shown in Table II.
 One can see that the results are actually independent of the method used to
 convert the energy difference to the alpha half-life.
 In a further analysis we use the recipee given in \cite{Roy}.

 In Fig. \ref{Q110}, we have shown calculated $Q_{\alpha}$ values for
 Darmsadtium isotopes.
 In the $g.s.\rightarrowtail g.s.$ transition, a semi-magic gap predicted at N=162
 is clearly visible. A very different behavior can be seen for
  $10^{-}_{\pi} \rightarrowtail 10^{-}_{\pi}$ and $10^{-}_{\nu} \rightarrowtail
 10^{-}_{\nu}$ transitions. The first show an even deeper minimum in
  $Q_{\alpha}$ than the one for the g.s. and it would signal then an
 extra stable nuclear state; the second show an opposite
 situation. Unfortunately, the excitation energy of both types of states is
 quite high and rather does not suggest their isomeric character.

 The Fig. \ref{Q110} confirms the proton character of states with delayed
 alpha-decay, and the effect of the proton configuration on
  the total isometric half-life in transitions of the type
 $20^{+}_{\pi \nu} \rightarrowtail 20^{+}_{\pi \nu}$ for $^{266-270}$Ds.
 A very interesting situation can be observed in $^{264}$Ds where
 $Q_{\alpha}$ values for two-neutron and two-proton configurations
 are similar and both lie significantly below the energy
   for the $g.s.\rightarrowtail g.s.$ transition.
  These transitions will be therefore much slower (~$10^{-4}$) than that
    between the ground states.
\begin{figure}[h]
\centerline{\includegraphics[scale=0.85]{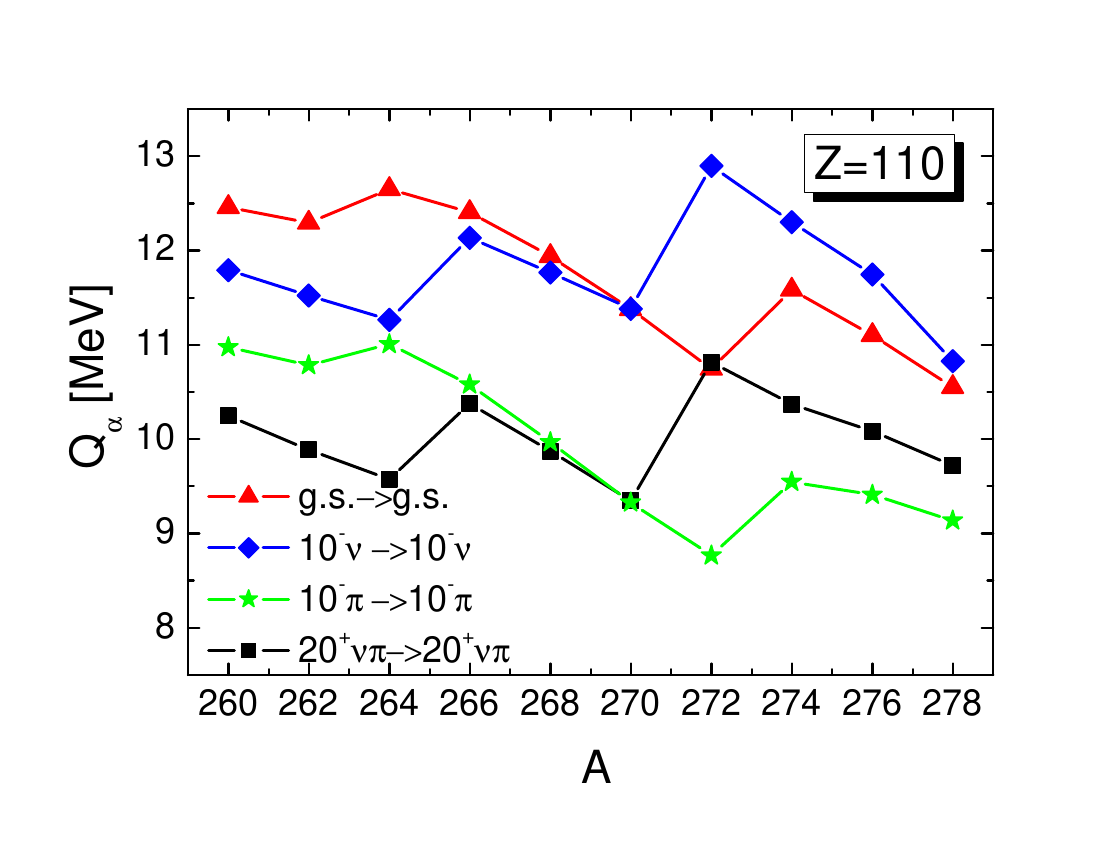}}
 \caption{{\protect  $Q_{\alpha}$ energies for transitions between states with the same structure for Ds.}}
\label{Q110}
\end{figure}
\begin{figure}[h]
\centerline{\includegraphics[scale=0.85]{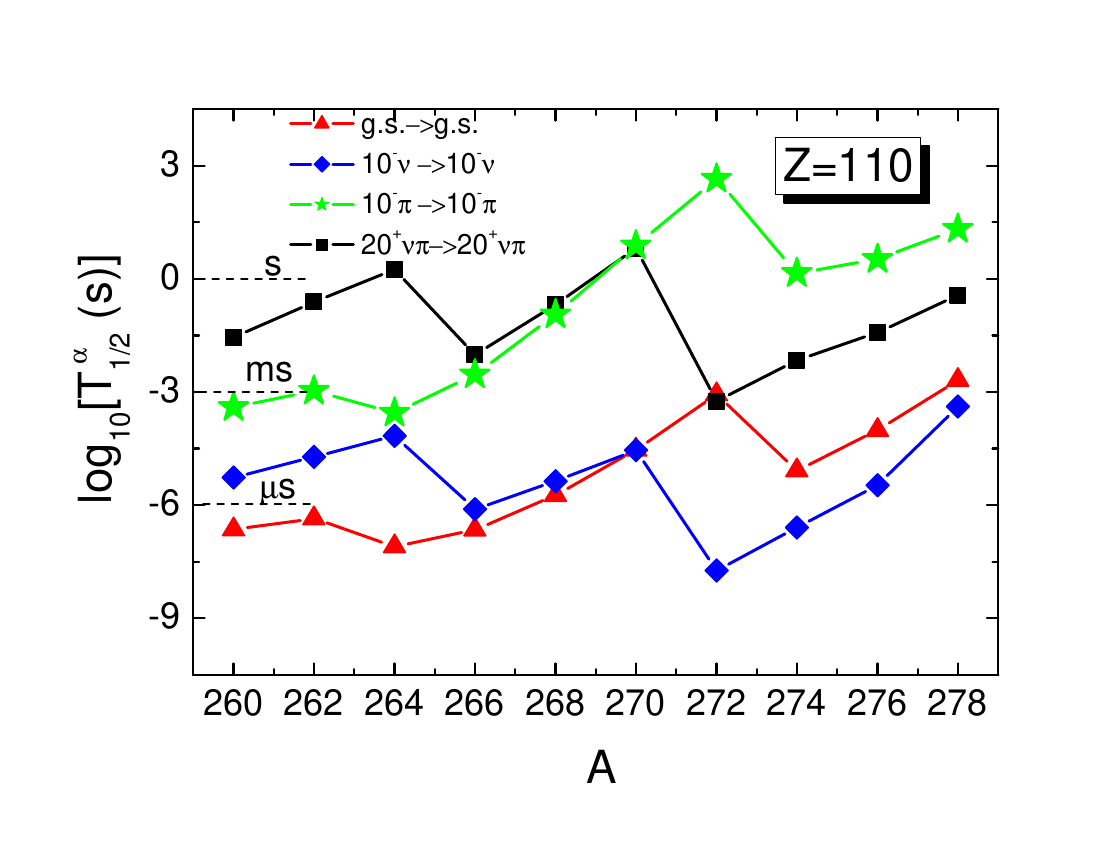}}
 \caption{{\protect $Log_{10}T_{\alpha}$ for Ds}}
\label{T110}
\end{figure}
 The alpha half-lives corresponding to the discussed transition energies
  are shown in Fig \ref{T110}.
 One can see, for example, that according to our calculation, a two-neutron
 isomeric state in $^{270}$Ds does not live longer than the ground state, as
  it was found experimentally for the isomer.
Recently, Clark and Rudolph \cite{Clark} obtained the proper half-life (3.9 ms)
 of the isomer in the frame of the superfluid tunnelling model (by assuming
 two-neutron q.p. character of the isomeric state, taking the experimentally
 measured $Q_{\alpha}$ value and arbitrarily weakening the pairing gap in
 the isomeric $10^{-}_{\nu}$ state).
One can see in Fig. \ref{T110} that in $^{264}$Ds, as well as in $^{270}$Ds,
the g.s. to g.s. decay occurs with a half life of $0.1 \mu s$, while for the
 4 q.p. state, the predicted half life is of the order of seconds.
This allows to think about chemical studies of isomeric states instead of
 ground states.

The final results - the hindrance factors for Ds isotopic chain -
 are shown in Fig. \ref{HF110}.
One can see that, contrary to the suggestion from the experimental paper on
 $^{270}$Ds \cite{Hof},
 the decay of the two-neutron quasi-particle ($10^{-}_{\nu}$) state is not at
 all hindered, while the decay of the proton two quasi-particle state
 ($10^{-}_{\pi}$) is strongly forbidden: $Log_{10}[HF]_{Q} = 5.42$.
The most prominent hindrance of the alpha decay among the four quasi-particle
 ($K^{\pi} = 20^{+}$) states - $10^{8}$ - is predicted for $^{264}$Ds.
 However, due to the short g.s. half life, the total half life for this
 particular isomer will be practically on the same level as for $^{270}$Ds.
 Thus, in addition to two nuclei, $^{266}$Ds and $^{268}$Ds, proposed by Liu
 \cite{Liu1} as candidates for  long-lived configurations,
 even more pronounced effect of an enhanced stability against alpha decay
 from the $K^{\pi} = 20^{+}$ state is predicted in $^{264}$Ds and $^{270}$Ds.
\begin{figure}[h]
\centerline{\includegraphics[scale=0.85]{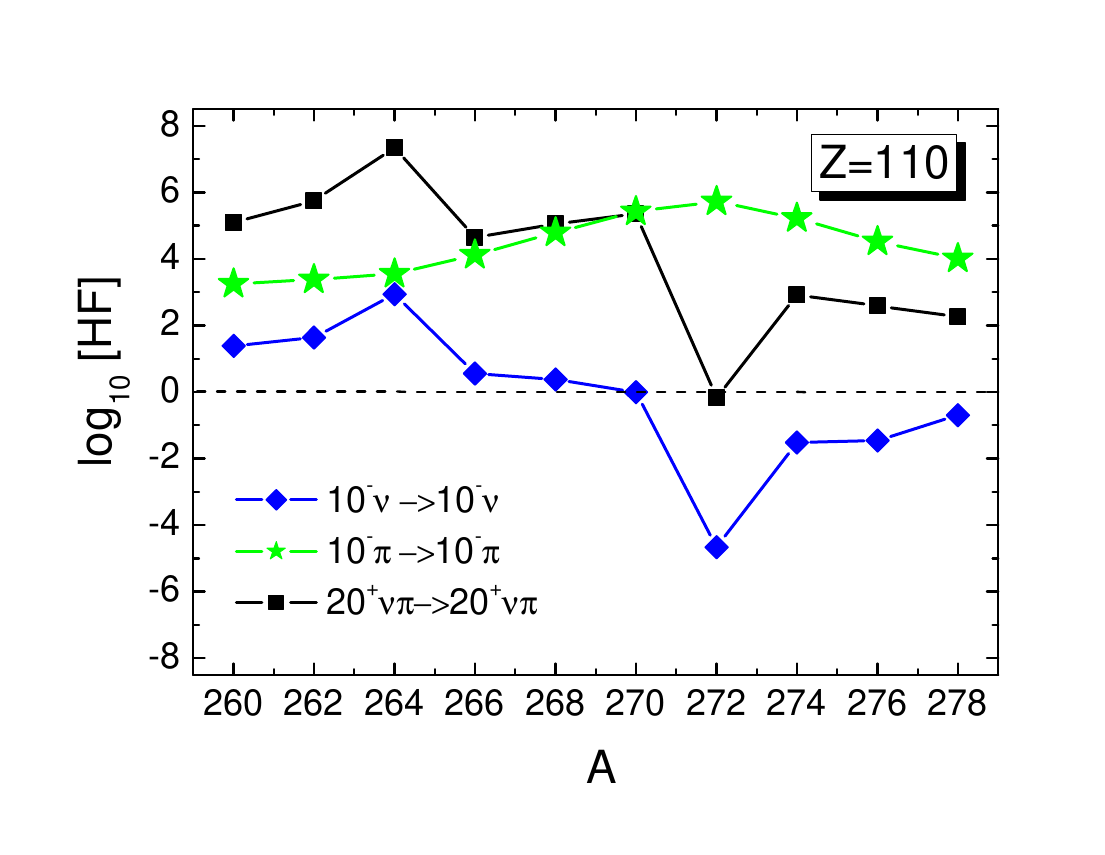}}
 \caption{{\protect  Hindrance factors in alpha decay of Ds. }}
\label{HF110}
\end{figure}
Our calculations also indicate that the decay of the 4q.p. state will not be
 hindered at all in $^{268}$Cn and $^{270}$Cn, contrary to what was suggested
 in the same paper \cite{Liu1}.
Admittedly, for these nuclei, the energies of these high-K states
 are sufficiently low to make them candidates for high-K isomers, but
 as follows from the discussion of excitations in $^{270}$Cn
 (Fig. \ref{rysschem}), we do not expect any hindrance here.
 On the contrary, a decay from the isomeric states should be faster
 than from the ground states.
 This happens in all Cn nuclei in which the excitation energy of the
 2 or 4 q.p. state is low, as shown in Fig. \ref{HF112}.
 From this figure one can see the only one exception, namely $^{266}$Cn,
 in which we expect some hindrance of the alpha decay from the isomer
 built on the neutron excitation.
However, the predicted hindrance is not very large ($ \simeq 10^{3}$).
\begin{figure}[h]
\centerline{\includegraphics[scale=0.85]{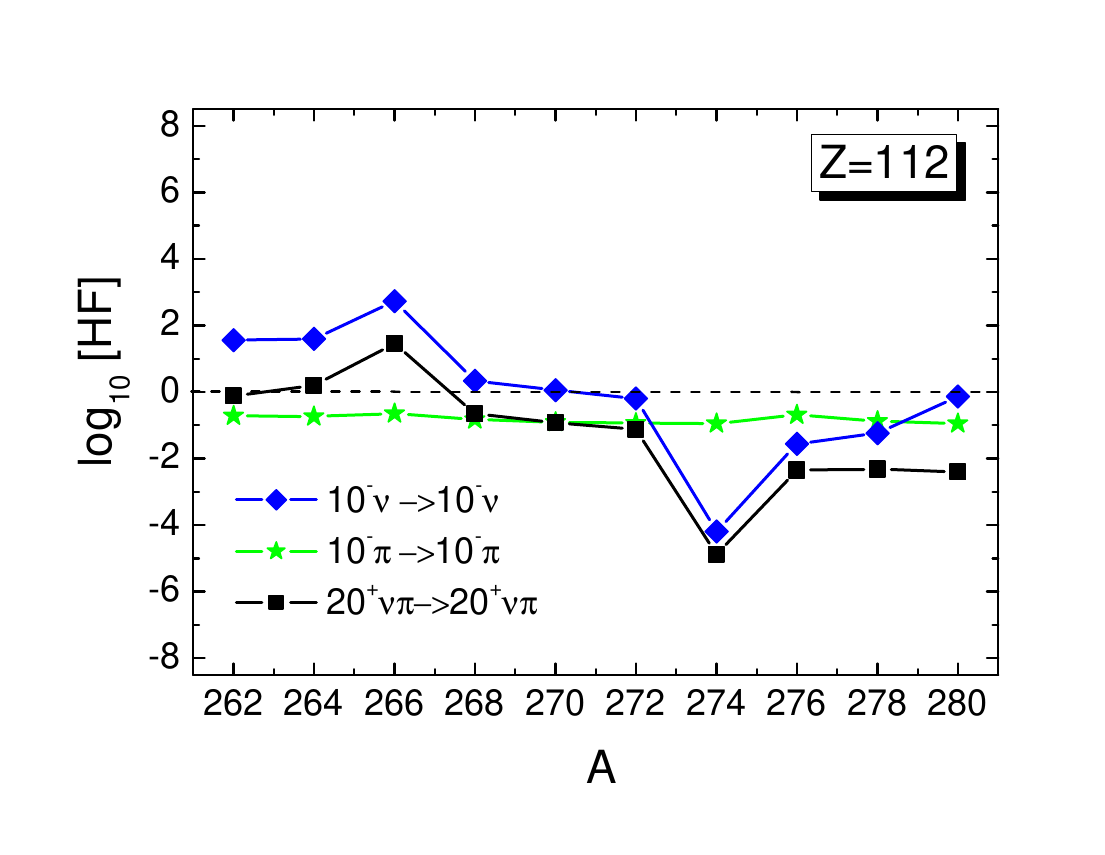}}
 \caption{{\protect  Hindrance factors in alpha decay of Cn. }}
\label{HF112}
\end{figure}

\section{CONCLUSIONS}

To summarize our investigations concerning the nature and behaviour of multi-quasiparticle states
in the superheavy nuclei in the context of their alpha-decay process:

i) We have found a quite strong hindrance against alpha decay for
 four quasi-particle states: $K^{\pi} = 20^{+}$ and/or $19^{+}$.
This, together with their relatively low excitation suggests a possibility
 that they could be isomers with an extra stability
  - five and more orders of magnitute longer-lived than the ground states.
 This would mean that chemical studies of such exotic high-K sates would be
 more likely than for quite unstable ground states.
 Among all tested nuclei, the best candidates for long-lived high-K isomers
 are predicted in $^{264-270}$Ds.

ii) Except a moderate (about 3 orders of magnitude) alpha-decay hindrance in
 $^{266}$Cn for a 2 q.p. neutron state, there are no more
candidates for an enhanced stability against alpha decay in Cn nuclei.

iii) Contrary to what has been recognized so far, our analysis indicates
 that the alpha-decay hindrance results mainly from the proton 2q.p.
  component.

\section*{ACKNOWLEDGEMENTS}

M.K. and J.S. were co-financed by the National Science Centre under Contract No. UMO-2013/08/M/ST2/00257  (LEA COPIGAL).


\begin{thebibliography}{99}


\bibitem{Walker}
P. M. Walker and G. D. Dracoulis, {\it Nature (London)}, {\bf 399} 35, (1999).

\bibitem{Xu}
F.R. Xu, E. G. Zhao, R. Wyss, and P. M. Walker {\it Phys. Rev Lett.} {\bf 92} 252501, (2004).







\bibitem{Hof}
S. Hofmann et. al.,{\it Eur. Phys. J. A}, {\bf 10}, 5 (2001).

\bibitem{Her}
R.-D. Herzberg et al., {\it Nature (London)} {\bf 442}, 896 (2006).

\bibitem{Tan}
S. K. Tandel et al., {\it Phys. Rev. Lett.} {\bf 97}, 082502 (2006).

\bibitem{Kon}
F. G. Kondev et al., {\it in Proceedings of the International Conference on Nuclear Data for Science and Technology,
Nice, France}, 2007, edited by O. Bersillon (EDP Sciences, Les Ulis Cedex, France, 2008).

\bibitem{Hes}
F. P. Heßberger et al., {\it Eur. Phys. J. A} {\bf 43}, 55 (2010).

\bibitem{Cla}
R. M. Clark et al., {\it Phys. Lett. B}  {\bf 690}, 19 (2010).

\bibitem{SDO}
P. Jachimowicz, M. Kowal, J. Skalski,   {\it Phys. Rev. C} {\bf 83}, 054302,  (2011).

\bibitem{Kisomers}
P. Jachimowicz, M. Kowal, and J. Skalski, {\it Phys. Rev. C},  {\bf92}, 044306 (2015).

\bibitem{Liu1}
H.L. Liu, P.M. Walker, and F. R. Xu, {\it Phys. Rev. C}, {\bf 89}, 044304 (2014).

\bibitem{Patyk1}
Z. Patyk, A. Sobiczewski, {\it Nuclear Phys. A} 533, 132 (1991).

\bibitem{Patyk2}
Z. Patyk, A. Sobiczewski, {\it Phys. Lett. B}  256,  307 (1991).



\bibitem{Ack} D. Ackermann, {\it Nucl.
Phys A} {\bf 944} (2015).

\bibitem{Asai}
M. Asai, F.P. Heßberger, A. Lopez-Martens, {\it Nucl.
Phys A} {\bf 944} (2015).

\bibitem{Drac}
G. D. Dracoulis et. al.  {\it Rep. Prog. Phys.}, {\bf 79} 076301 (2016).

\bibitem{walkerPS}
P. M. Walker and F. R. Xu, {\it Phys. Scr.}, {\bf 91} 013010 (2016).

\bibitem{kon2}
F. G. Kondev, G. D. Dracoulis, T. Kibedi, {\it Atomic Data and Nuclear Data Tables}, 103-104, (2015).

\bibitem{David}
H. M. David et al., {\it Phys. Rev. Lett.}  {\bf 115}, 132502 (2015).

\bibitem{WS}
 S.~\'Cwiok, J.~Dudek, W.~Nazarewicz, J.~Skalski and T.~Werner, {\it Comput. Phys. Commun.} {\bf 46}, 379 (1987).

\bibitem{KN}
H.~J.~Krappe, J.~R.~Nix and A.~J.~Sierk, {\it Phys. Rev. C} {\bf20}, 992 (1979).

\bibitem{WSpar}
  I.~Muntian, Z.~Patyk and A.~Sobiczewski, {\it Acta Phys. Pol. B} {\bf 32}, 691 (2001).

\bibitem{Kow}
 M. Kowal, P. Jachimowicz, A. Sobiczewski, {\it Phys. Rev. C} {\bf 82}, 014303 (2010).

\bibitem{kowskal}
 M. Kowal, J. Skalski  {\it Phys. Rev. C} {\bf 82}, 054303 (2010).

\bibitem{IIbarriers}
P. Jachimowicz, M. Kowal, J. Skalski  {\it Phys. Rev. C}, {\bf85}, 034305 (2012).

\bibitem{IIIbarriers1}
M. Kowal, J. Skalski  {\it Phys. Rev. C}, {\bf85}, 061302(R) (2012).

\bibitem{IIIbarriers2}
P. Jachimowicz, M. Kowal, J. Skalski,  {\it Phys. Rev. C}, 044308 (2013).

\bibitem{archive}
 M. Kowal, P. Jachimowicz, J. Skalski,  arXiv:1203.5013 (2012).

\bibitem{jach2017}
P. Jachimowicz, M. Kowal, and J. Skalski, {\it Phys. Rev. C},  {\bf95}, 034329 (2017).

\bibitem{jachbar2017}
P. Jachimowicz, M. Kowal, and J. Skalski, {\it Phys. Rev. C},  {\bf95}, 014303 (2017).

\bibitem{BroSkal}
W. Brodzinski and J. Skalski, {\it Phys. Rev. C} {\bf88}, 044307 (2013).

\bibitem{Liu2011} H. L. Liu, F. R. Xu, P. M. Walker, and C. A. Bertulani,
 {\it Phys. Rev. C} {\bf 83}, 011303(R) (2011).

\bibitem{Hi}
S.Hilaire, M. Girod, {\it AIP Conference Proceedings} 1012, 359 (2008).

\bibitem{Delion}
D. S. Delion, R. J. Liota, R. Wyss, {\it Phys. Rev. C}, {\bf76} 044301 (2007).

\bibitem{Karamian}
S. A. Karamian, J. J. Carroll, S. Iliev, and S. P. Tretyakova
 {\it Phys. Rev. C} {\bf 75}, 057301 (2007).

\bibitem{Jach2014}
P. Jachimowicz, M. Kowal, J. Skalski  {\it Phys. Rev. C}, {\bf89 } 024304 (2014).

\bibitem{dub}
$http://nrv.jinr.ru/nrv/webnrv/alpha_decay/index.php$

\bibitem{Roy}
G. Royer, {\it Nucl. Phys.A} {\bf 848}, 279 (2010).

\bibitem{park}
A. Parkhomenko, and A. Sobiczewski, Acta Phys. Pol. B 36, 3095 (2005).

\bibitem{Clark}
R. M. Clark and D. Rudolph, {\it Phys. Rev. C} {\bf97}, 024333 (2018).


\end{thebibliography}
\end{document}